\documentclass{elsart}
\usepackage[english]{babel}
\usepackage{latexsym,amssymb,amsmath}

\usepackage{natbib}
\usepackage{graphicx}
\usepackage{amssymb}

\begin{document}
\begin{frontmatter}

\title{Spectral properties of backscattered solar Ly-$\alpha$ radiation in the heliosphere: a theoretical search of the heliospheric boundaries effects}
\author[MSU,IKI]{Olga A. Katushkina\corauthref{cor}},
\corauth[cor]{Corresponding author.}
\ead{okat@iki.rssi.ru}
\author[MSU,IKI,IPMeh]{Vladislav V. Izmodenov}
\ead{izmod@iki.rssi.ru}

\address[MSU]{Department of Aeromechanics and Gas Dynamics, Faculty of Mathematics and Mechanics, Lomonosov Moscow State University,
Moscow 119899, Russia}
\address[IKI]{Space Research Institute (IKI), Moscow 117997, Russia}
\address[IPMeh]{Institute for Problems in Mechanics Russian Academy of Sciences, Moscow, Russia}

\begin{abstract}
We explore imprints of the solar wind interaction with the local interstellar medium on spectral properties of the backscattered solar Ly-$\alpha$ radiation.
We employ newly developed effective model for the interstellar H atom velocity distribution function in the heliosphere (Katushkina \& Izmodenov 2010, 2011). This model takes into account both global effects of H atom perturbations at the heliospheric boundaries and local (i.e. within 10-20 AU from the Sun) effects of the solar ionization, charge exchange, solar gravitation and radiation pressure.

 Backscattered solar Ly-$\alpha$ profiles and their zero, first and second moments were computed at 1 AU for the anti-solar directions of line-of-sight. Then the moments are compared with the moments calculated by using more simplified one-component and two-component hot models. The comparison shows that the Ly-$\alpha$ intensities calculated in the frame of new model are somewhat smaller than those calculated by means of the simplified models. Small differences in the first moment (i.e. line-shift) are also observed in the downwind direction.

The largest difference between new model and hot models is seen in the second moment of the backscattered Ly-$\alpha$ profile. This moment is also called as line-of-sight ``temperature'' or line-width of the backscattered spectra. The spectral width depends on the width of the H atom velocity distribution function within few AU from the Sun. We show that the line-widths calculated on the basis of the our model is qualitatively different from those of simplified models.

It is also shown that minimum of the line-width at 60$^{\circ}$ from upwind that was observed by SOHO/SWAN \citep{costa99, quem_etal06} is absent in the frame of our model. Such a minimum is expected from simple consideration of two populations (primary and secondary) of interstellar H atoms and although new model includes two populations of H atoms we do not see this minimum in our results. Instead of this minimum, our model predicts a small local maximum of the line-width at $\sim$150$^{\circ}$ from upwind. We explain these effects by non-Maxwellian behavior of the velocity distribution function of H atoms at entrance to the heliosphere. More specifically, difference in the kinetic temperatures $T_{x}$ and $T_{z}$ plays a key role. Physical phenomena which can help to get the observed minimum in the frame of our model are discussed.

\end{abstract}

\begin{keyword}
heliosphere \sep interstellar hydrogen \sep hot model \sep solar Ly-$\alpha$ radiation
\PACS 96.50.Xy \sep 96.50.Zc \sep 98.70.Vc

\end{keyword}

\end{frontmatter}

\parindent=0.5 cm

\section{Introduction}

\begin{figure}
\begin{center}
\includegraphics[scale=0.8]{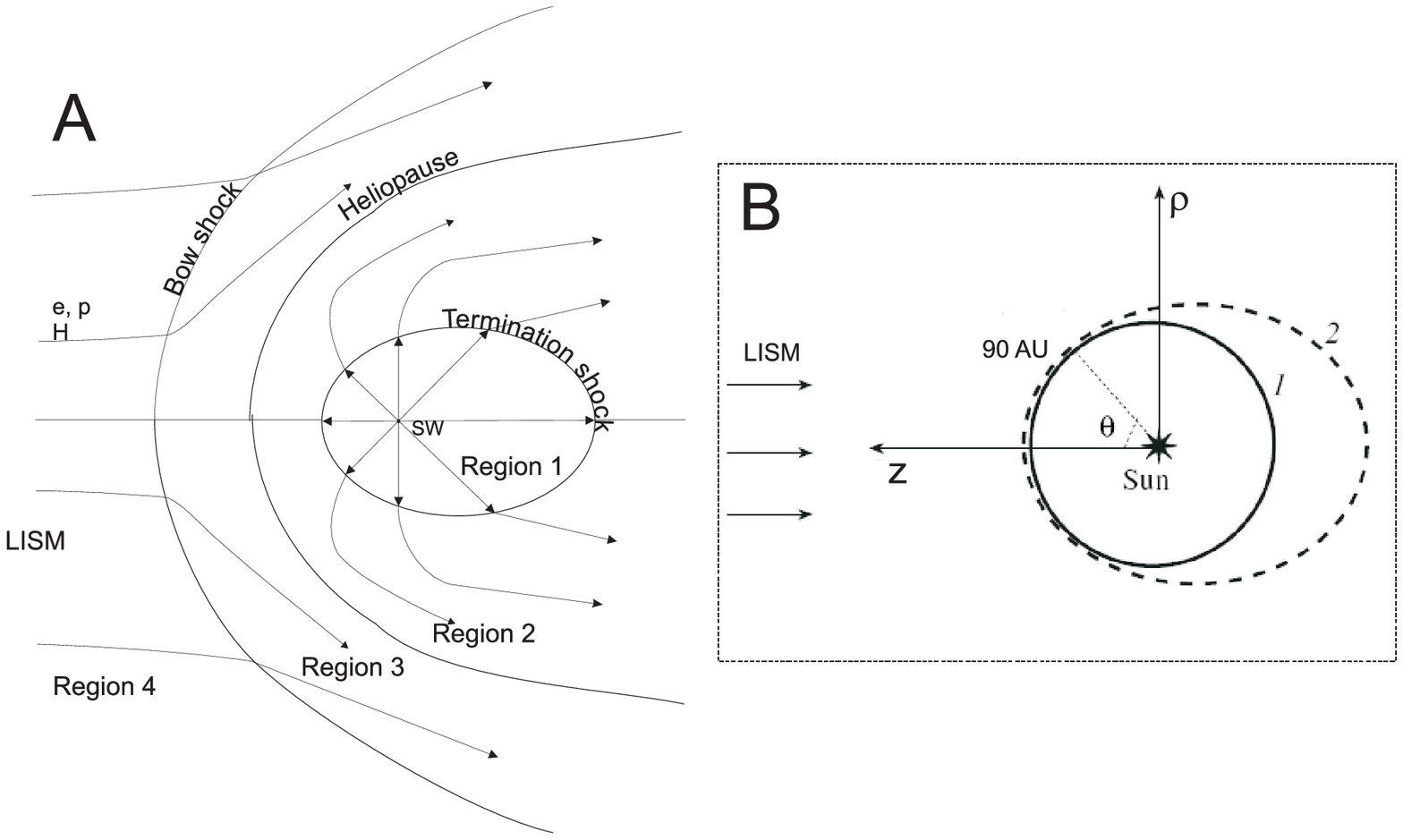}
\end{center}
\caption{A: Qualitative picture of the heliospheric interface; B: coordinate system used in the models: $z$ is the axis of symmetry directed toward the upwind direction, $\rho$ is an axis in cylindrical coordinate system that is perpendicular to the upwind direction; curve~1 is the outer boundary in our model; curve~2 is a schematic representation of the heliospheric termination shock.}
\label{interface-sc}
\end{figure}

Our solar system is moving though the Circum-Heliospheric Interstellar Medium (CHISM).
The solar wind interacts with interstellar plasma and complex structure
of the interaction region is formed due to relative Sun/CHISM motion (see Fig.\ref{interface-sc}~A). CHISM is a weakly (20-30~\%) ionized plasma and mainly consists of hydrogen. The mean free path of the interstellar hydrogen atoms (H atoms) is larger or comparable with the characteristic distance of the SW/CHISM interaction region. Therefore, H atoms penetrate deeply to the heliosphere, where properties of H atom component can be measured directly on indirectly. Due to their large mean free path in the heliosphere a kinetic approach should be used to describe H atom component theoretically.

 The first self-consistent kinetic-gasdynamical model of the SW/CHISM interaction has been developed by \citet{bm93} (Baranov-Malama model hereafter). It was shown
 that the SW/CHISM interaction region (that is often called in literature as the heliospheric interface) consists of four regions with different
 plasma properties. These regions are separated by three discontinuity surfaces, namely, the termination shock and bow shock are boundaries of the supersonic
 solar wind and interstellar medium flows, accordingly, and the heliopause is a tangential surface separated charged components of the two plasma flows.

One of the results obtained by \citet{bm93} and previously by \citet{baranov_etal91} is the hydrogen wall that was discovered experimentally by \citet{linsky_wood96} (see also Wood et al., 2007a, 2007b for recent developments). The hydrogen wall is an increase of H atom density in the vicinity of the heliopause due to production of newly created H atoms by charge exchange. These secondary atoms have properties of plasma in the vicinity of the heliopause. The interstellar plasma in the vicinity of the heliopause is heated and decelerated, and, therefore, the mean velocity of the secondary H atom population is smaller and the temperature is larger as compared with the velocity and temperature of the original (or primary) H atom population that consists of interstellar atoms, which entering into the heliopshere without being charge exchanged in the heliospheric interface. A mixture of the primary and secondary interstellar H atom populations penetrate to the heliosphere. Inside the heliosphere in region between the heliopause and the termination shock (this region is often called as the inner heliosheath) new atoms are also created during charge-exchange. These atoms are called as the energetic neutral atoms (ENAs) and now they are measured by Interstellar Boundary Explorer spacecraft \citep{mccomas_etal09}. ENAs have a small number density in the heliosphere as compared with interstellar atoms, that is why ENAs are not important for backscattered Ly-$\alpha$ radiation.

It was recognized since 1970s that the solar Ly-$\alpha$  photons are effectively backscattered
by the interstellar H atoms. Spectral properties of the backscattered Ly-$\alpha$ radiation depend on the velocity and spatial distribution of the interstellar hydrogen inside the heliosphere. Since the distribution of H atoms is disturbed in the heliospheric interface, then the backscattered Ly-$\alpha$ may, in principle, serve as the source of information on the heliospheric interface structure. For example, \citet{costa99} analyzed SOHO/SWAN data and showed that the temperature of the interstellar hydrogen at large distances from the Sun (at 50 AU where local effects of solar gravitation, radiation and ionization are negligible) is significantly higher then temperature of the CHISM. The bulk velocity of the hydrogen flow obtained by this analysis is about 21 km/s that is $\sim$5 km/s smaller than the relative SW/CHISM velocity. These differences in the properties of the hydrogen gas as compared with the properties of the CHISM are explained by existence of the secondary interstellar H atom population in the heliopshere.

The effect of the heliospheric interface is also seen in the measured \citep{lallement05} deflection of the interstellar H atom flow direction as compared with the direction of the interstellar helium. In contrast to the hydrogen the charge-exchange cross-section of helium ($He + H^{+} \leftrightarrow He^{+} + H$) is small, that is why it is usually assumed that the interstellar helium stay practically undisturbed during its motion from CHISM to the heliosphere. It means that the velocity vector of the helium flow in the vicinity of the Sun refers to the value and to direction of the interstellar gas flow. In turn, hydrogen atom flow is affected by charge-exchange with protons in the SW/CHISM interaction region. The deflection of hydrogen atoms inside the heliosphere from the direction of SW/CHISM relative motion is most probably due to distortion of the global heliospheric interface structure caused by the interstellar magnetic field that leads to asymmetry of interstellar plasma flow in the heliospheric interface and this asymmetry transfers to H atoms through the charge-exchange. This fact opened a possibility to estimate magnitude and direction of the interstellar magnetic field \citep[e.g.][]{izmod_alex05}.

Another manifestation of the heliospheric interface was found during analyzes of the second moment of the backscattered Ly-$\alpha$ radiation spectra. This moment corresponds to the width of the spectra and depends on the width of the H atom velocity distribution. Being expressed in the units of temperature the line-width is often called as the line-of-sight (or l-o-s) temperature. \citet{costa99} and \citet{quem_etal06} derived the l-o-s temperature of the backscattered Ly-$\alpha$ from SOHO/SWAN data and presented the temperature as a function of the angle $\theta$ between the line-of-sight direction and the direction toward the CHISM flow (i.e. the upwind direction). It was shown that the temperature in the directions of $\theta$ = 50-60$^{\circ}$ is smaller by 1500-2000~K than the temperature toward upwind. This minimum was explained by \citet{costa99} and \citet{quem_etal06} qualitatively by presence of the primary and secondary populations of the interstellar hydrogen atoms.

The goal of this study is to theoretically explore the imprints of the heliospheric interface on spectral properties of the backscattered Ly-$\alpha$ radiation and separate these effects from the local effects (i.e. the effects that act within 10-20 AU of the Sun) of the solar ionization, charge exchange, solar gravitation and radiation pressure. We have computed line profiles and their moments of the backscattered Ly-$\alpha$ spectra based on newly developed advanced kinetic model of the hydrogen distribution in the heliosphere (Katushkina \& Izmodenov, 2010, 2011). This model combines comparative simplicity of the classical hot model with ability to account for both the perturbations of the hydrogen distribution function in the SW/CHISM interaction region (later it will be mentioned as effects of the heliospheric interface) and local effects which are significant only in the vicinity of the Sun. In order to separate the heliospheric boundary effects from the local effects we compare new model results with the results of the simplified hot models, which describe local effects in similar way as the advanced model, but do not take into account the effects of the heliospheric boundaries.

\section{Modeling of the hydrogen distribution inside the heliosphere} \label{H_distr}

To model spatial and velocity distribution of the interstellar hydrogen in the heliosphere we solve a kinetic equation (see, e.g. Katushkina \& Izmodenov, 2010) for the distribution function $f(\textbf{r},\textbf{w},t)$ of the hydrogen atoms inside the sphere with radius of 90 AU around the Sun (see Fig.~\ref{interface-sc}~B). The processes of the photoionization, charge-exchange, solar gravitation and solar radiation pressure are taken into account. It is assumed in the model that the rates of charge exchange $\beta_{ex}$ and photoionization $\beta_{ph}$ decrease with distance from the Sun as $1/r^{2}$. We use following values of ionization rates at the Earth orbit: $\beta_{ph,E}=1.16\times 10^{-7}$ s$^{-1}$, $\beta_{ex,E}=4.8\times 10^{-7}$ s$^{-1} $. To calculate the charge-exchange rate at the Earth orbit we use averaged values of the solar wind number density (6~cm$^{-3}$) and velocity (440~km/s) known from measurements, and charge-exchange cross-section from \citet{lindsay_steb05}.
The solar gravitational force and the force of the radiation pressure act in opposite directions and both are proportional to $1/r^{2}$, where $r$ is a distance to the Sun. Dimensionless parameter $\mu$ defines a balance between the solar gravitation and radiation pressure. In this work we assume that $\mu$ is equal to 1.258.

In this paper we consider three models which differ each from other by boundary conditions at 90 AU.
{\bf Model 1} is the one-component hot model that implies the Maxwellian velocity distribution function at the outer boundary. For the specific calculations performed in the frame of this model we use parameters of the Maxwellian distribution (i.e. number density, bulk velocity and temperature) calculated at 90 AU in the upwind direction in the frame of the Baranov-Malama model with parameters of chosen to be the same as in \citet{katush_izmod10}. These values at 90 AU are: n$_{H}$ = 0.54 $\cdot n_{H,LISM}$,  V$_{H,z}$ = -0.79$\cdot V_{H,LISM}$, T$_{H}$ = 15 000~K, where n$_{H,LISM}$=0.18~cm$^{-3}$, V$_{H,LISM}$=26.4~km/s. These parameters correspond to the mixture of the primary and secondary interstellar atoms. Note, the effects of the heliospheric interface are taken in this model at zero level by taking parameters in the Maxwellian distribution different from the CHISM parameters.

{\bf Model 2}  is so-called two-component hot model \citep[e.g.][]{bzowski_etal08}. In this case model~1 is employed separately for the primary and secondary populations of interstellar atoms. The velocity distribution functions are assumed to be Maxwellian for each of the populations and parameters of the distribution functions were calculated in the frame of Baranov-Malama model. These parameters for the primary interstellar atoms are: n$_{H,prim}$ = 0.22 $\cdot n_{H,LISM}$,  V$_{z,prim}$ = -1.06$\cdot V_{H,LISM}$, T$_{H,prim}$ = 6 840~K, and for the secondary interstellar atoms: n$_{H,second}$ = 0.32 $\cdot n_{H,LISM}$,  V$_{z,second}$ = -0.63$\cdot V_{H,LISM}$, T$_{H,second}$ = 18 126~K.

As it was shown in \citet{katush_izmod10}, model 2 leads to significant discrepancies in the distribution of H atoms inside the
heliosphere as compared with the self-consistent Baranov-Malama model. This fact is connected with non-Maxwellian behavior of the velocity distribution
function of hydrogen atoms after their crossing the heliospheric interface \citep{izmod_etal01, izmod01}.
%\textbf{V stat'e 2010 goda ne rassmatrivalas' model 1...}

In {\bf model 3} we use the boundary conditions allowing to get the distribution of H atom inside the heliosphere in very good agreement with the distribution obtained in the frame of Baranov-Malama model.
For the primary interstellar atoms the distribution function at 90 AU is chosen as 3D normal distribution. In this case all zero, first and second moments of the velocity distribution function are taken into account \citep[see][]{katush_izmod10} and calculated in the frame of Baranov-Malama model.
However, for the population of the secondary interstellar atoms such approach does not give the full agreement with the Baranov-Malama model because the distribution function of this population is asymmetric with respect to its maximum and it has nonzero third moments which are neglected in the 3D normal distribution. That is why for the secondary interstellar atoms \citet{katush_izmod11} have used the velocity distribution function of at 90 AU calculated by Monte-Carlo in the frame of the Baranov-Malama model. It was shown that 15-20 millions of unsplitted \citep{malama91} trajectories in the Monte-Carlo code is enough in order to get acceptable accuracy of calculations. In this paper we follow \citet{katush_izmod11} approach for the primary interstellar atoms and use the numerically calculated velocity distribution function as the boundary conditions at 90 AU for the secondary interstellar atoms.

It is important to note here that for the stationary and axisymmetric case model 3 provides the same distributions within 90 AU from the Sun as the Baranov-Malama model. However, computationally our model is more effective and allows to calculate the velocity distribution function everywhere inside the heliosphere with required precision. Moreover, our model allows to study local (i.e. within the heliosphere) time-dependent and 3D phenomena very effectively.

\section{Modeling of the backscattered solar Ly-$\alpha$ profiles}

The backscattered Ly-$\alpha$ profiles I(\textbf{r},$\nu$,$\mathbf \Omega$) were computed for anti-solar radial directions ($\mathbf \Omega$) for an observer located at 1 AU. Here \textbf{r} is a position of observer, $\nu$ is a frequency of the backscattered radiation, $\mathbf \Omega$ is the line-of-sight.
We use ``self-absorption'' approximation \citep{quem00}. In this approximation only once scattered photons are considered (i.e. multiple scattering of photons are neglected), and absorption of the photons between the Sun and the scattering point is neglected. Despite this simplified ``self-absorption'' approach gives similar results as compared with a full radiative transfer model. \citet{quem_izmod02} have shown that for the line-width of the backscattered profile at 1 AU the difference between full radiative transfer model and the self-absorption model is less than 15 \% for the upwind direction and becomes smaller as the line-of-sight moves from upwind to downwind. In the downwind direction the two approaches give practically the same line-widths. The simplified self-absorption approach is sufficient for the purposes of this paper. As it will be seen the effects of the heliospheric interface are essentially larger than the difference between the full radiative transfer and self-absorption models.

The radiative transfer equation for I(\textbf{r},$\nu$,$\mathbf \Omega$) can be written as follows:
\begin{equation}
 \mathbf \Omega \cdot \nabla I(\textbf{r},\nu,\mathbf \Omega) =-\sigma_{\nu}(\textbf{r},\nu)n_{H}(\textbf{r})I(\textbf{r},\nu,\mathbf \Omega) + n_{H}(\textbf{r})j(\textbf{r},\nu,-\mathbf \Omega)\label{transfer}.
\end{equation}
Here $n_{H}(\textbf{r})$ is the number density of hydrogen atoms, $\sigma_{\nu}(\textbf{r},\nu)$ is the differential scattering cross-section that is
  proportional to the projection of the hydrogen distribution function on the line-of-sight, j($\textbf{r}$,$\nu$,-$\mathbf{\Omega}$) is the atomic emission
  coefficient which measures the number of photons emitted by a hydrogen atom per second per frequency unit and per solid angle. Note, that
  scattered photons travel in the direction opposite to the line-of-sight direction i.e. in $-\mathbf \Omega$. The first term in the right hand of the equation (\ref{transfer}) is a loss term due to absorption of the scattered photons and the second term is a local source of the emission due to scattering process.

  The equation~(\ref{transfer}) has a classical solution in case of the self-absorption approximation:
\begin{equation}
  I(\textbf{r},\nu,\mathbf \Omega) = \frac{4\,\pi}{10^{6}}\,\int_{0}^{\infty}n_{H}(\textbf{r}+s\mathbf \Omega)\, j(\textbf{r}+s\mathbf \Omega,\nu,-\mathbf \Omega) \, e^{-\tau_{\nu}(\textbf{r}+s\mathbf \Omega,\textbf{r})}ds\label{intens},
\end{equation}
where, $s$ is a coordinate along the line-of-sight, $\tau_{\nu}(\textbf{r}',\textbf{r})$ is the optical thickness for scattered photons with frequency $\nu$ going from
the scattered point $\textbf{r}'=\textbf{r}+s\mathbf \Omega$ to observer at point $\textbf{r}$. Ly-$\alpha$ profile is measured here in Rayleigh. In this solution only once scattered photons are taken into account, and absorption of primary solar photons between the Sun and scattering point is neglected.
Atomic emission coefficient can be represented as:
  \begin{equation}
     j(\textbf{r}',\nu,-\mathbf \Omega) = \phi(\omega) \, F_{S}(\textbf{r}',\nu_{p})\,\sigma_{\nu}(\textbf{r}',\nu)\, \label{g}.
 \end{equation}
  Here $\phi(\omega)$ is a scattering phase function that expresses the relation between direction of propagation of the photon before and after scattering
  \citep{brandt_cham59}. $F_{S}(\textbf{r}',\nu_{p})$ is a flux of primary (solar) Ly-$\alpha$ photons with the frequency of $\nu_{p}$ at point $\textbf{r}'$. We use the solar Ly-$\alpha$ spectra \citep{lemair_etal98} to calculate solar Ly-$\alpha$ flux at the Earth orbit with defined frequency.
  In that case when line-of-sight is radial, there is a simple relation between frequency of the primary solar photon $\nu_{p}$ and frequency of
  the backscattered photon $\nu$: $\nu_{p}=2\cdot\nu_{0}-\nu$, $\nu_{0}$ is the frequency at line center.
  Thus, if we know the hydrogen velocity distribution function in the whole heliosphere, we can calculate the profiles of the backscattered Ly-$\alpha$ radiation.

   Since we consider the two different populations of interstellar H atoms inside the heliosphere it is possible to calculate profiles of the radiation
  scattered by each populations separately. To do this we consider photons that were scattered by the primary and secondary interstellar atoms independently. Optical thickness is calculated for the mixture of the primary and secondary atoms, since a photon scattered, for example, on the primary interstellar atom may then be absorbed by both the primary or secondary atoms.

 We calculate the following moments of the backscattered radiation profiles:

   \[
     I_{los}(\textbf{r},\mathbf \Omega) = \int_{0}^{\infty}I(\textbf{r},\nu,\mathbf \Omega)\,d\nu \mbox{ -- intensity in Rayleigh;}
   \]
   \[
     V_{los}(\textbf{r},\mathbf \Omega) = \frac{\int_{0}^{\infty}u(\nu)\,I(\textbf{r},\nu,\mathbf \Omega)\,d\nu}{I_{los}(\textbf{r},\mathbf \Omega)} \mbox{ -- line-shift expressed in [km/s];}
   \]
   \[
     T_{los}(\textbf{r},\mathbf \Omega) = \frac{m_{H}}{k_{b}}\,\frac{\int_{0}^{\infty}(u(\nu)-V_{los}(\textbf{r},\mathbf \Omega))^{2}\,I(\textbf{r},\nu,\mathbf \Omega)\,d\nu}{I_{los}(\textbf{r},\mathbf \Omega)}\mbox{ -- line-width expressed in [K].}
   \]

  Here, $u(\nu)=c\,(\nu/\nu_{0}-1)$, $m_{H}$ is the mass of a hydrogen atom, $k_{b}$ is the Boltzmann constant. The line-shift of the spectra is often called as the line-of sight velocity, and the line-width is also called as the line-of-sight (or apparent) temperature.
  These integral characteristics of the backscattered Ly-$\alpha$ profile reflect the properties of the velocity distribution function of H atoms inside the heliosphere, but they do not coincide and should not be mixed up with the bulk velocity and temperature of hydrogen atoms far away from the Sun.

\section{Results of the modeling}

We computed the backscattered Ly-$\alpha$ profiles and their moments by making use of three described above models of the hydrogen
distribution inside the heliosphere. All computations were performed for the anti-solar directions. For the 2D axisymmetric problem considered here each line of sight is characterized by one angle $\theta$ that is counted from z-axis (see Fig.\ref{interface-sc}~B).

\begin{figure}
\begin{center}
\includegraphics[scale=0.7]{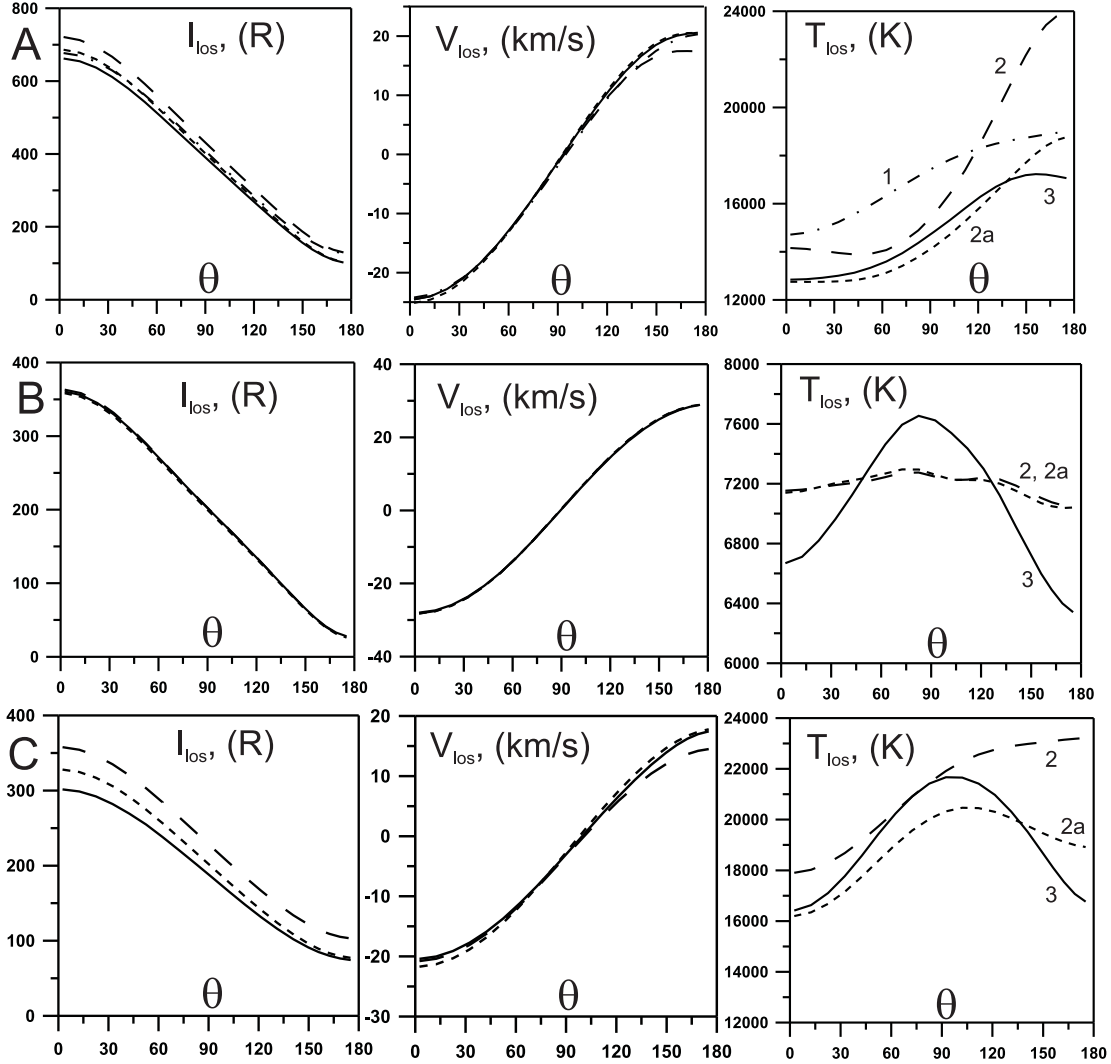}
\end{center}
\caption{ Intensities (left column), line-shifts (center-column) and line-widths (right column) of the backscattered Ly-$\alpha$ radiation at 1 AU as functions of the line-of-sight angle $\theta$. Plots A (top row) are for the total radiation scattered by both primary and secondary interstellar atoms. Plots B (middle row) are for photons that were scattered by the primary interstellar atoms. Plots C (bottom row) correspond to the radiation scattered by the secondary interstellar atoms. Different curves correspond to three models of the hydrogen distributions in the heliosphere: (1) is the one-component hot model (model 1); (2) is the two-component hot model (model 2) ; (3) is our model (model 3) that takes into account effects of the heliospheric interface; additional curves marked as 2a in plots A,B,C correspond to model of H atoms described in this section, it is model 2a that is equal to model 2 plus
$\theta$-dependence of hydrogen parameters at the outer boundary.
}
\label{Lya_mom}
\end{figure}

Fig.~\ref{Lya_mom} shows the intensities (left column), line-shifts (center-column) and line-widths (right column) of the backscattered Ly-$\alpha$ radiation at 1 AU as functions of the line-of-sight angle $\theta$. For models 2 and 3 the profiles of the photons scattered by the primary and secondary interstellar H atoms were computed separately (plots B, C in Fig.~\ref{Lya_mom}). The total backscattered profile was calculated as well (plot A). For model~1 that has only one component of hydrogen we compute only characteristics of the total Ly-$\alpha$ radiation.

For the total radiation (plots A) models 1 and 2 lead to systematic increase of the intensities as compared with the intensities calculated for model 3. Comparison of models 2 and 3 for the primary and secondary populations (plots B and C) shows that the increase in intensities is due to the secondary H atom component (compare left-columns in the plots A and C). Models 2 and 3 are nicely agreed for the primary H atom component .

Now, we will explore the main reason of the different results produced by models 2 and 3. The models are only different in the boundary condition. We can identify two differences of the boundary conditions. The first is the dependence of the boundary conditions on the angle $\theta$ in model 3 that does not exist in model 2, since model 2 assumes constant parameters at the boundary. The second difference is the non-Maxwellian velocity distribution function at 90 AU in model 3, while in model 2 the velocity distribution is Maxwellian. To explore which of the two identified reasons is mainly responsible for the differences in the spectral properties of Ly-$\alpha$ we perform additional model calculation (model 2a). In model 2a we assume that velocity distribution functions at the outer boundary of 90 AU are Maxwellain for both the primary and the secondary populations of H atoms. However, the parameters of the Maxwellians (i.e. number density, bulk velocity and temperature) are functions of the angle $\theta$ that were calculated in the frame of Baranov-Malama model. Therefore, model 2a allows to separate the effects of non-uniform flow of H atoms at 90 AU after they passed the heliospheric interface from the kinetic effects connected with non-Maxwellian features of the velocity distribution function.

Results of model 2a are also shown in Fig.~\ref{Lya_mom}~A. It is seen that model 2a and model 3 produce very close results in the backscattered Ly-$\alpha$ intensities, although there is a difference of 20 rayleigh for the upwind direction. It means that the main difference of models 2 and 3 is due to non-uniform flow of H atoms at 90 AU in model 3. However, still there is a smaller difference due to non-Maxwellian behavior of the velocity distribution function at 90 AU (which is taken into account only in model 3).

\begin{figure}
\begin{center}
\includegraphics[scale=0.5]{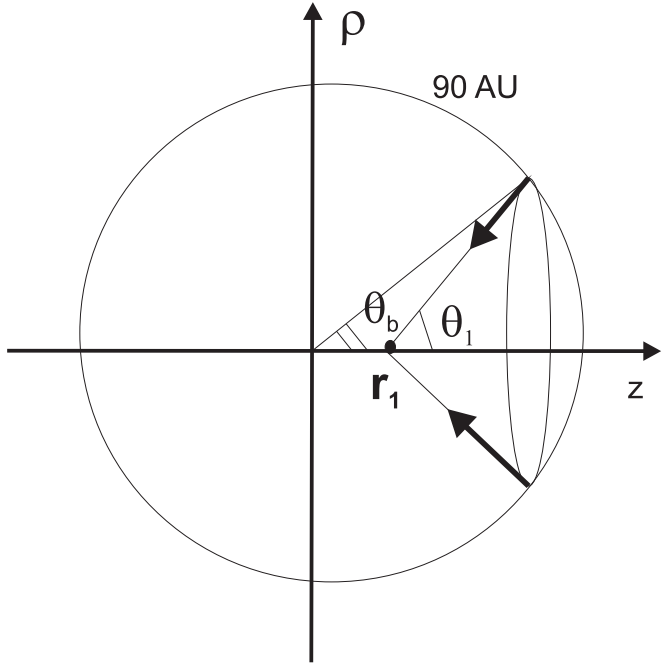}
\end{center}
\caption{ Schematic picture of penetration of the hydrogen atoms from the outer boundary to the vicinity of the Sun. Point 1 ($\textbf{r}_{1}$) is located
in the upwind direction at 2 AU from the Sun. In the case of straight atom's trajectories we consider only those atoms that arrive to point 1 from
the outer boundary at $\theta_{b}$. Angle $\theta_{1}$ can be determined if we know $r_{1}$ and $\theta_{b}$.
}
\label{konus}
\end{figure}

\begin{figure}
\begin{center}
\includegraphics[scale=0.7]{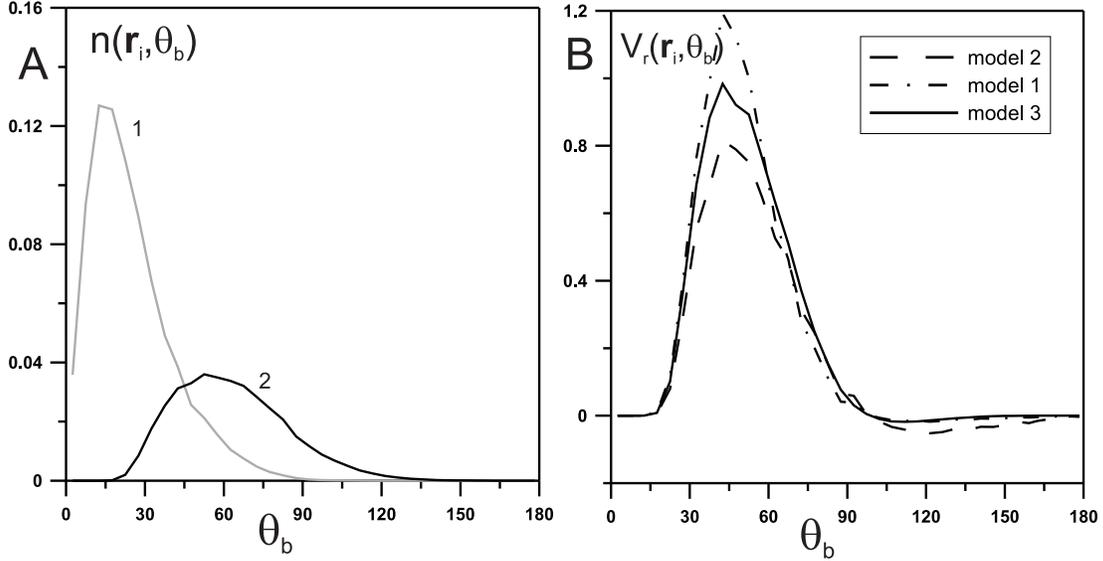}
\end{center}
\caption{ Contributions to the total number density (A) and radial velocity (B) of atoms arrived from various directions. Curve~1 in plot A corresponds to point~1 ($r=2$ AU, $\theta$=0$^\circ$), and curve~2 corresponds  point~2 ($r=7$ AU, $\theta$=180$^{\circ}$).
Contribution to the radial velocity was calculated only for point~2. Different curves in plot B correspond to different models of hydrogen distribution: solid line is model 3, dashed line is model 2 and dashed-dot line is model 1. All calculations are performed for the mixture of primary and secondary interstellar atoms.
}
\label{n_vr_L}
\end{figure}

At first view, it is not evident why the angle dependence of the H atom parameters at 90 AU would strongly influence the backscattered Ly-$\alpha$ emission measured at 1 AU. It is especially so, because the main emissivity region for the backscattered Ly-$\alpha$ radiation at 1 AU is located approximately at 2 AU for the upwind line-of-sight and at 7 AU for downwind. From the simple (naive) consideration one could expect that the most of H atoms in the regions close to the Sun would arrive from upwind. However, \cite{lallem_bert90} have shown (in the case of the hot model) that the most of the atoms arrive into the vicinity of the Sun not from upwind. The same is true for our model. To illustrate this we calculate function $n(\textbf{r}_i,\theta_{b})$ for two points (i=1,2) inside the maximum emissivity region. Point 1 is located in the upwind direction at 2~AU from the Sun. Point 2 is located in the downwind direction at 7~AU from the Sun. The function $n(\textbf{r}_i,\theta_{b})$ represents the contribution to the total number density at a given point (point 1 for curve 1 in Fig.~\ref{n_vr_L}~A
and point 2 for curve 2) from those particles that cross the outer boundary of 90 AU at $\theta = \theta_{b}$.
 This function $n(\textbf{r}_i,\theta_b)$ is defined as follows (for certainty we present all formulas only for point 1):
\begin{equation}
  n(\textbf{r}_{1},\theta_{b})= \int_{\Omega_{1}} f(\textbf{r}_{1},\textbf{w}_{1})d\textbf{w}_{1} \label{n_theta}
\end{equation}
Here $f$ is the velocity distribution function of H atom. The integration is performed over those $\textbf{w}_{1}$ that correspond to the trajectories crossing the outer sphere of 90 AU at $\theta = \theta_{b}$. Here for simplicity we consider a balance between the solar gravitation and radiation pressure ($\mu=1$). This means that all atom's trajectories are straight lines. In this case subspace $\Omega_{1}$ is a cone with apex
angle equal to $\theta_{1}$ as it is seen from Fig.~\ref{konus}. Let us introduce a spherical coordinate system in the velocity space. It means that we describe velocity vector $\textbf{w}$ by its module $\tilde{w}$ and two spherical angles $\tilde{\theta}$ and $\tilde{\varphi}$, i.e. orthogonal coordinates of vector $\textbf{w}$ can be represented
as follows:
\begin{equation}
 \left\{
       \begin{aligned}
        w_{x}&=\tilde{w}\cdot sin(\tilde{\theta})\cdot cos(\tilde{\varphi})\\
        w_{y}&=\tilde{w}\cdot sin(\tilde{\theta})\cdot sin(\tilde{\varphi})\\
        w_{z}&=-\tilde{w}\cdot cos(\tilde{\theta}) \\
        \end{aligned}
 \right.
\end{equation}

In spherical coordinates $d\textbf{w}_{1}=\tilde{w}_{1}^{2} \, sin(\tilde{\theta}) \, d\tilde{w}_{1} \, d\tilde{\theta} \, d\tilde{\varphi}$. For velocities
from the subspace $\Omega_{1}$: $\tilde{\theta}=\theta_{1}=const$ for chosen value of $\theta_{b}$ and integration over $\tilde{\theta}$ is not needed. Hence, for our case integral~(\ref{n_theta}) can be rewritten in the following form:
\begin{equation}
 n(\textbf{r}_{1},\theta_{b})=\int_{0}^{+\infty} \int_{0}^{2\pi} f(\textbf{r}_{1}, \tilde{w}_{1}, \tilde{\theta}=\theta_{1}, \tilde{\varphi}) \, \tilde{w}_{1}^{2} \, sin(\theta_{1}) \, d\tilde{w}_{1} \, d\tilde{\varphi}. \label{n_2}
\end{equation}

Now, in the case of $\mu=1$:
\[
f(\textbf{r}_{1}, \tilde{w}_{1}, \theta_{1}, \tilde{\varphi})=f_{b}(\theta_{b}, \tilde{w}_{1}, \theta_{1}, \tilde{\varphi})\cdot \exp(-A(\textbf{r}_{1}, \theta_{b}, \tilde{w}_{1})),
\]
where $f_{b}$ is a corresponding velocity distribution function at 90 AU, $A$ is the loss of hydrogen atoms along its trajectory from the outer boundary to point 1 due to ionization processes and $\theta_{1}=\theta_{1}(\theta_{b})$. $f_{b}$ does not depend on angle $\tilde{\varphi}$ due to axial symmetry of the boundary conditions. Therefore expression~(\ref{n_2}) can be represented as:
\begin{equation*}
 \begin{aligned}
 n(\textbf{r}_{1},\theta_{b})&=\int_{0}^{+\infty} \int_{0}^{2\pi} f_{b}(\theta_{b},\tilde{w}_{1},\theta_{1}(\theta_{b})) \, \exp(-A(\textbf{r}_{1}, \theta_{b}, \tilde{w}_{1})) \, \tilde{w}_{1}^{2} \, sin(\tilde{\theta}_{1}) \, d\tilde{w}_{1} \, d\tilde{\varphi} = \\
 &=2\pi \, sin(\theta_{1}) \cdot \int_{0}^{+\infty} f_{b}(\theta_{b},\tilde{w}_{1},\theta_{1}(\theta_{b})) \, \exp(-A(\textbf{r}_{1}, \theta_{b}, \tilde{w}_{1})) \, \tilde{w}_{1}^{2} \, d\tilde{w}_{1} = \\
 &=2\pi \, sin(\theta_{1}) \cdot g(\textbf{r}_{1},\theta_{b}).
 \end{aligned}
\end{equation*}

Function $f_{b}(\theta_{b},\tilde{w}_{1})$ decreases with $\theta_{b}$ for each given value of $\tilde{w}_{1}$, because at 90 AU $V_{z,H}\gg V_{\rho,H}$ and distribution function has a maximum in the
upwind direction. Loss-function $A(\textbf{r}_{1}, \theta_{b}, \tilde{w}_{1})$ increases with $\theta_{b}$ for point 1 because length of the trajectory has a minimal value in upwind and $\exp(-A)$ decreases with $\theta_{b}$. Hence, function $g(\textbf{r}_{1},\theta_{b})$ decreases with $\theta_{b}$. And $sin (\theta_{1})$ increases with $\theta_b$ for $\theta_b\in[0,\pi/2]$. Therefore $n(\textbf{r}_{1},\theta_{b})$ that is a product of $sin (\theta_{1})$ and $g(\textbf{r}_{1},\theta_{b})$ should have a maximum for some $\theta_{b}$ between 0$^\circ$ and 90$^\circ$.

Fig.~\ref{n_vr_L}~A shows $n(\textbf{r}_i,\theta_b)$ calculated numerically for point 1 in upwind (curve~1) and for point 2 in downwind (curve~2). It is seen that the
most part of the interstellar atoms arrive to the point~1  from $\theta\approx15^{\circ}$ and to the point~2  from $\theta\approx55^{\circ}$. Therefore, contrary to the simple consideration the most of H atoms arrive to the vicinity of the Sun not exactly from upwind.

Now, it is clear that the excess in the backscattered Ly-$\alpha$ intensity in models 1 and 2 as compared with the model 3 is connected with larger number densities of H atoms inside the heliosphere in models 1 and 2 due to non varying with $\theta$ boundary conditions while in model 3 the number density at 90 AU decreases with $\theta$.

It also becomes clear why the intensities (and line-shifts) of the radiation scattered by the primary interstellar atoms are in agreement for all models (Fig.~\ref{Lya_mom}~B). This is because on the one hand, angle-dependence of hydrogen parameters at 90 AU for the primary interstellar atoms is weaker than for the secondary interstellar atoms, and on the other hand, distribution function of primary interstellar atoms is closer to Maxwellian than distribution function of secondary interstellar atoms.

From the middle column of Fig.~\ref{Lya_mom} (A,C) it is seen that there are noticeable differences in the line-shifts of models 2 and 3 in the downwind direction. The differences are seen for the total radiation as well as for the radiation scattered by the secondary interstellar atoms.
At the same time there is almost no discrepancies in line-shifts of model 1 and model 3. In spite of model 1 is the simplest model without any effects of the heliospheric interface, and one could expect the difference.

In order to understand these results we computed the contribution to the total radial velocity of hydrogen at point~2 (7 AU in downwind) from the particles that reach this point from the different directions. Similar to $n(\textbf{r}_{i},\theta_{b})$ contribution to the radial velocity can be calculated as follows:
\[
  V_{r}(\textbf{r}_{i},\theta_{b})=\frac{1}{\int_{-\infty}^{+\infty}f(\textbf{r}_{i},\textbf{w}_{i})d\textbf{w}_{i}}\cdot \int_{\Omega_{i}}f(\textbf{r}_{i},\textbf{w}_{i})\, w_{i,r} \,d\textbf{w}_{i}
\]
Fig.~\ref{n_vr_L}~B shows $V_{r}(\textbf{r}_{i},\theta_{b})$, calculated at point~2 for different models.
It is seen that the maximum of $V_{r}(\textbf{r}_{i},\theta_{b})$ is located approximately at $\theta$=45$^{\circ}$ for all models. However, model 2 (dashed curve in Fig. 3~B) has a non-negligible contribution of the negative values of $V_{r}(\textbf{r}_{i},\theta_{b})$ for $\theta>100^{\circ}$. The negative values of $V_{r}(\textbf{r}_{i},\theta_{b})$ are due to relatively hotter secondary interstellar atoms in the downwind hemisphere (due to large temperature of secondary interstellar atoms and absence of decrease of the number density from upwind to downwind in model 2), which can reach point~2 from large values of angle $\theta$. This makes the total line-shift of model 2 in the downwind region smaller as compared with models 1 and 3. Contributions of negative $V_{r}(\textbf{r}_{i},\theta_{b})$ in model 1 and model 3 are smaller than in model 2 but due to different reasons. In model 3 the contribution of the particles with negative $V_{r}(\textbf{r}_{i},\theta_{b})$ are significantly reduced because of small number density of such particles at the outer boundary that follows from the self-consistent model results. As for model 1 we do not see this effect due to relatively smaller temperature of the mixture of primary and secondary interstellar atoms as compared with the temperature of the secondary interstellar atoms which exist in model 2.

Now let us consider the line-width of the Ly-$\alpha$ (right column in Fig.~\ref{Lya_mom}). Plot A demonstrates that there are essential qualitative differences in the line-widths calculated on the basis of the three models. Model 1 shows a monotonic increase of the line-width with the angle $\theta$. Model 2 shows a
 minimum of the line-width at $\theta \sim$60$^{\circ}$. Model 3 shows a small local maximum at $\theta \sim$150$^{\circ}$. These results demonstrate that the line-width of the backscattered Ly-$\alpha$ profiles is very sensitive to the properties of hydrogen distribution at the entrance to the heliosphere. In the next section we will focus on understanding of these qualitative differences and will show that the main causes of the difference are in kinetic non-Maxwellian nature of the hydrogen velocity distribution function at 90 AU. Note also, that for the line-widths the simplified model 2a results are close to model 3 in the upwind hemisphere. However, large discrepancies in the downwind region still take place.

 Now we will consider the differences in the line-widths of the backscattered spectra calculated separately for the primary and secondary populations of H atoms  (right columns in Fig.~\ref{Lya_mom}~B,C). For both populations we see the large maxima of the line-widths at $\theta \sim$ 90$^{\circ}$ in the results of model 3. Such strong maxima do not exist for model 2 and for model 2a. The latter means that the maxima are not due to $\theta$-dependence of hydrogen parameters at 90 AU but due to non-Maxwellian velocity distribution function at the outer boundary.
 Namely, as it was shown by \citet{baranov_etal98} (see also, Izmodenov et al., 2001, and Fig.~3~e,f in Katushkina and Izmodenov, 2010) components T$_{z}$ and T$_{\rho}$ of the kinetic temperatures calculated from the velocity distribution function (i.e. T$_{z}(\textbf{r})\sim \int f(\textbf{r},\textbf{w})\cdot (V_{z}(\textbf{r})-w_{z})^{2}d\textbf{w}$ and T$_{\rho}(\textbf{r})\sim \int f(\textbf{r},\textbf{w})\cdot (V_{\rho}(\textbf{r})-w_{\rho})^{2}d\textbf{w}$)
are essentially different. In other words the mean thermal velocities of the H atoms are different in different directions.
Moreover this difference between T$_{z}$ and T$_{\rho}$ temperatures
increases approaching to the Sun due to local effects. The large maxima of the line-widths at $\theta= 90^{\circ}$ (Fig.~\ref{Lya_mom}~B,C) are connected with the
changes of the radial kinetic temperature T$_{r}$ of H atoms at 90 AU as a function of $\theta$. For example, when observer looks toward upwind (i.e.
$\theta=0^{\circ}$) T$_{r}$=T$_{z}$, for the line-of-sight of $\theta=90^{\circ}$ T$_{r}$=T$_{\rho}$ and for line-of-sight of $\theta=180^{\circ}$  T$_{r}$=T$_{z}$ again. In model 3 at the outer boundary T$_{\rho}$ is larger than T$_{z}$ for each of the interstellar populations of H atoms (that follows from Baranov-Malama model results). This leads to the maxima of T$_{r}$ at $\theta=90^{\circ}$ and that are reflected in the maxima of the line-widths for photons scattered by each population of interstellar H atoms separately.

\begin{table}[!h]
 \vspace{6mm}
 \centering
  \caption[]{The comparison of intensities, line-shifts and line-widths of the backscattered Ly-$\alpha$ radiation for three considered models of the hydrogen distribution inside the heliosphere (description of the models see in text)}\label{tab1}
\vspace{5mm}\begin{tabular}{l|c|c|c|c|c|c}
\hline $\theta=0^{\circ}$ & $I_{los} (R)$ & $V_{los}$ (km/s) & $T_{los}$ (K) & $\frac{|I_{los}-I_{los,3}|}{I_{los,3}}\cdot100\%$ & $\frac{|V_{los}-V_{los,3}|}{V_{los,3}}\cdot100\%$ & $\frac{|T_{los}-T_{los,3}|}{T_{los,3}}\cdot100\%$\\
\hline $\mbox{model~1}$ & 677 & -24.2 & 14 710 & 2.3 & 1.2 & 14.5 \\
$\mbox{model~2}$ & 720 & -24.5 & 14 158 & 8.7 & 0.0 & 10.3 \\
$\mbox{model~3}$ & 662 & -24.5 & 12 841 & 0.0 & 0.0 & 0.0 \\
\hline $\theta=90^{\circ}$ & $I_{los} (R)$ & $V_{los}$ (km/s) & $T_{los}$ (K) & $\frac{|I_{los}-I_{los,3}|}{I_{los,3}}\cdot100\%$ & $\frac{|V_{los}-V_{los,3}|}{V_{los,3}}\cdot100\%$ & $\frac{|T_{los}-T_{los,3}|}{T_{los,3}}\cdot100\%$ \\
\hline $\mbox{model~1}$ & 398 & -0.9 & 17 505 & 4.7 & 200.0 & 18.0 \\
$\mbox{model~2}$ & 421 & -0.6 & 15 561 & 10.8 & 100.0 & 4.9\\
$\mbox{model~3}$ & 380 & -0.3 & 14 834 & 0.0 & 0.0 & 0.0 \\
\hline $\theta=180^{\circ}$ & $I_{los} (R)$ & $V_{los}$ (km/s) & $T_{los}$ (K) & $\frac{|I_{los}-I_{los,3}|}{I_{los,3}}\cdot100\%$ & $\frac{|V_{los}-V_{los,3}|}{V_{los,3}}\cdot100\%$ & $\frac{|T_{los}-T_{los,3}|}{T_{los,3}}\cdot100\%$ \\
\hline $\mbox{model~1}$ & 127 & 20.3 & 18 984 & 24.5 & 1.4 & 11.3 \\
$\mbox{model~2}$ & 129 & 17.4 & 23 946 & 26.5 & 15.5 & 40.4 \\
$\mbox{model~3}$ & 102 & 20.6 & 17 060 & 0.0 & 0.0 & 0.0 \\
\hline
\end{tabular}
\end{table}

Finally, to qualify the differences in the Ly-$\alpha$ intensities, line-shifts and line-widths calculated by making use of models 1-3 we present Table~\ref{tab1} that shows differences (absolute and relative) between the model results for the upwind ($\theta=0^{\circ}$), crosswind ($\theta=90^{\circ}$) and downwind ($\theta=180^{\circ}$) directions. It is seen that one-component hot model 1 gives from 2~\% (in upwind) to 24~\% (in downwind) differences with model 3 in intensities and about 10-18~\% differences in the line-widths. Two-component hot model 2 leads to 8-27~\%
discrepancies with model 3 in the intensities, and from 10~\% (in upwind) to 40~\% (in downwind) discrepancies in the line-widths.
The differences in the line-shifts of models 1 and 2 relative to model 3 are large especially for the crosswind direction. However, they are connected with the small values of the line-shifts and most probably can not be detected experimentally.

In the next section we discuss nature of the dependence of the Ly-$\alpha$ line-widths on the angle $\theta$.

\section{Line-width of the backscattered Ly-$\alpha$ profiles as diagnostics of the heliospheric interface nature}

\citet{costa99} and \citet{quem_etal06} have analyzed spectral properties of the backscattered solar Ly-$\alpha$ radiation measured by SOHO/SWAN
in 1996-2003. The Ly-$\alpha$ line-width as function of the line-of-sight direction was studied. It was shown that there is a noticeable minimum of the line-width at $\theta=50^{\circ}-60^{\circ}$. This minimum was interpreted as the indication of the two (primary and secondary) components of the interstellar H atoms in the heliosphere and, therefore, as an evidence of the heliospheric interface. Indeed, the results obtained in the frame of
the one-component classical hot model show (dashed-dot curve on Fig. 2A) the monotonic increase of the line-width from upwind to downwind.
Existence of the two components, which have the bulk velocities shifted one with respect to other, can help to produce the minimum of the temperature in the directions close to the crosswind direction.

Models 2 and 3 considered in this paper have the two-components of H atoms at the outer boundary at 90 AU. Therefore, one can expect that the minimum of the line-widths should be obtained in these models. However, as it is seen from Fig.~\ref{Lya_mom}~A the minimum is seen only for model 2 and not seen for model 3.
Instead, for model 3 we see a small maximum of the temperature at $\theta=150^{\circ}$. In order to explain these features
we have performed series of additional test calculations.

Our goal is to understand and distinguish roles of different effects influencing the
line-width of the Ly-$\alpha$ profiles. To do that we studied the effects of all possible model parameters and established that the following factors influence the dependence of the line-width ($T_{los}$) on the angle $\theta$:

\begin{description}
   \item[1.] Ionization processes that change the parameters of H atoms near the Sun. From our test calculations we found that the solar gravitation and radiation pressure have smaller influence on the line-width than ionization.
   \item[2.] Existence of the two populations of the interstellar hydrogen atoms that are shifted in velocity space. This effect leads to appearance of the
   minimum of $T_{los}$ at $\theta=90^{\circ}$ as it was discussed before.
   \item[3.] Non-Maxwellian behavior of the velocity distribution function of the two populations at 90 AU. More precisely, the large difference between $T_{z}$ and $T_{\rho}$ "temperature" components ($z$ and $\rho$ are cylindrical coordinates).
   This effect takes place for model 3.
   \item[4.] Shape of the solar spectrum of the Ly-$\alpha$ radiation. It will be shown below that the shape of the solar spectrum has some small but interesting
   effect on the line-width as function of $\theta$.
 \end{description}

\begin{figure}
\begin{center}
\includegraphics[scale=0.7]{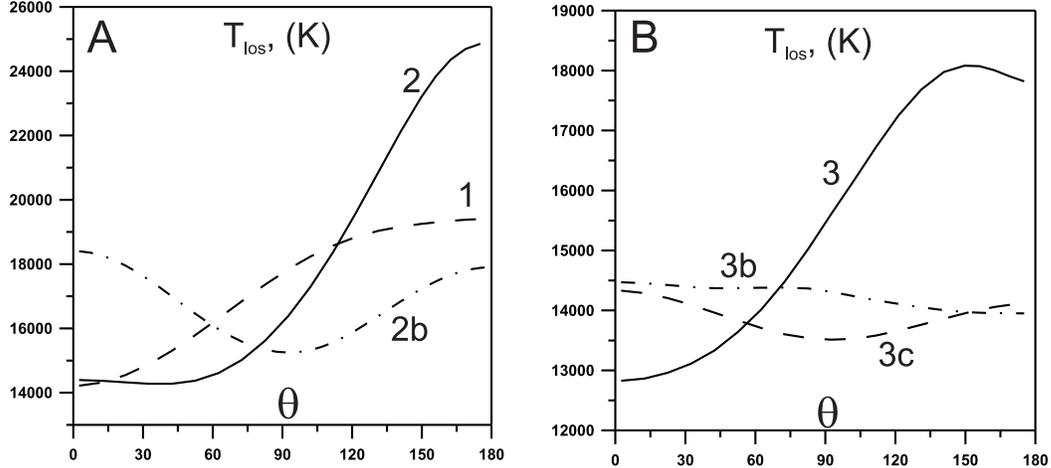}
\end{center}
\caption{ Results of test calculations of the Ly-$\alpha$ line-width as a function of the angle $\theta$.
Results based on the one-component hot model (model 1 curve 1) and the two-component hot model (model 2 curves 2 and 2b) are shown in
plot (A) and results based on model 3 (curves 3, 3b, 3c) are shown in plot (B).
Curves 2 and 2b correspond to calculations based on model 2 with and without ionization, respectively; curve 3 corresponds to model 3, and curve 3b and 3c
are the results of model 3 where the ionization was switched off; in all calculations except model 3c solar spectrum from \citet{lemair_etal98} was assumed, for model 3c the flat solar spectrum was employed. Also see Table~\ref{table_2}.
}
\label{T_tot}
\end{figure}

Fig.~\ref{T_tot} summaries the results of test-calculations (see, also, table~\ref{table_2} for model numeration) and explore the listed above effects. For all calculations of this section we assume the solar gravitation force to be in balance with the solar radiation, i.e. the parameter
$\mu=1$ and the trajectories of the H atoms are straight lines. To explore the effects of ionization in the vicinity of the Sun we performed test calculations with typical ionization rate of $\beta_{E}=5.9 \cdot 10^{-7}\, s^{-1}$ as well as the calculations with $\beta_{E}=0$ (models 2b, 3b, 3c).

\begin{table}[!h]
  \caption[]{Parameters of the test calculations}\label{table_2}
\begin{tabular}{p{2cm}||c}
\hline { \scriptsize curve's number on fig.~\ref{T_tot}} &  Model description \\
\hline 1  & one-component model (1) \\
\hline 2  & two-component model (2) \\
\hline 2b & two-component model (2) without ionization \\
\hline 3  & our model (3)  \\
\hline 3b & our model (3) without ionization  \\
\hline 3c & our model (3) without ionization; flat solar spectrum of the Ly-$\alpha$ radiation  \\
\hline

\end{tabular}

\end{table}

The effects of the local ionization on the function of $T_{los}(\theta)$ are clearly illustrated in the frame of the one-component hot model 1 (curve 1 in
Fig.~\ref{T_tot}~A) since neither two-component nor non-Maxwellian effects are taken into account in this model. The line-width increases monotonically from the upwind direction to downwind. Such a behavior of the line-width of the backscattered Ly-$\alpha$ profile reflects the properties of the H atom distribution since the ionization (due to so-called selection effect) leads to decrease of effective temperature of H atoms in upwind and increase of the temperature in the downwind direction (see for example Fig.~3.5 in \citet{izmod_issi06}).

Results of the two-component hot model 2b that does not account for ionization (curve 2b on Fig.~\ref{T_tot}) illustrate the effect of the two populations of the interstellar H atoms.
These two populations have different bulk velocities $V_{z}$ (at 90 AU in the upwind direction: $V_{z,primary}\approx-27$~km/s and $V_{z,secondary}\approx-16$~km/s) and rather small thermal velocities. Therefore in the upwind and downwind directions (where $V_{r}=\pm V_{z}$) the line-of-sight projections of the velocity distribution functions of the primary and secondary components are overlap each other in the velocity space only partially. In the crosswind direction $V_{r}=V_{\rho}\approx 0$ for both primary and secondary interstellar atoms. This means that in this direction the projections of the distribution functions on the radial line-of-sight overlap completely. That is why the radial temperature of the mixture of primary and secondary interstellar atoms is smaller in the crosswind direction than in the upwind and downwind directions. This minimum of the radial temperature of H atoms in crosswind is reflected in the Ly-$\alpha$ line-widths as it is clearly seen from model 2b.

Results of the two-component hot model (curve~2) where the ionization is taken into account combine both the increase of $T_{los}$ from upwind to downwind due to ionization effect and the local minimum in crosswind due to two populations of H atoms. Hence, a small minimum of $T_{los}$ at $\theta$=50-60$^{\circ}$ is seen in curve~2.

The line-width obtained by making use of model 3 is presented in Fig.~\ref{T_tot}~B. Note that this model takes into account
all considered effects of the heliospheric interface, namely: 1) two populations of interstellar H atoms, 2) $\theta$-dependence of the hydrogen parameters at 90 AU, 3) non-Maxwellian features of the hydrogen velocity distribution function at the entrance to the heliosphere. In particular, differences between kinetic temperature's components $T_{z}$ and $T_{\rho}$ play an important role.

Curve 3c corresponds to the model 3c that does not take into account ionization and in addition in model 3c
we assume that the solar Ly-$\alpha$ flux does not depend on frequency. Note that in all other models we use the shape of the solar spectrum from
\citet{lemair_etal98}.

It is seen that the effect of minimum in $T_{los}$ at $\theta=90^{\circ}$ (due to two populations) practically disappears in model 3c, but still is visible. This practical disappearance of the minimum is connected with the increase of the radial kinetic temperature T$_{r}$ with increase $\theta$ at 90 AU. The effect of T$_{r}$ increase with $\theta$ compensates the effect of the minimum of the radial hydrogen temperature in crosswind due to the two populations of H atoms. That is why the value of the minimum of the line-widths at $\theta=90^{\circ}$ is much smaller for model 3c as compared with model 2b.

It is interesting to note, that for model 3b (that is more realistic than model 3c) local minimum of $T_{los}$ at $\theta$ = 90$^{\circ}$ replaces
by small maximum. This effect is due to shape of the solar spectrum. Remember, that in model 3c we use flat solar spectrum while in model 3b we use
nonuniform solar spectra from \citet{lemair_etal98}.

%The line-width in crosswind for model 3b is larger (by about 1000 K) than for model 3c. This is the effect of the
%shape of the solar Ly-$\alpha$ spectrum. In realistic solar spectrum of model 3b the intensity of the solar radiation is the smallest at the line center. Along crosswind line-of-sight most part of H atoms has small radial component of velocity and it means that most part of solar photons that can be scattered by such atoms were emitted from the line center, while for upwind and crosswind line-of-sight most part os scattered photons have a doppler-shift relative to the line center. It leads to decrease of the
%maximum intensity of backscattered radiation in crosswind as compared for example with upwind. And it corresponds to increase of the line-width of
%spectra in crosswind.

\begin{figure}
\begin{center}
\includegraphics[scale=0.7]{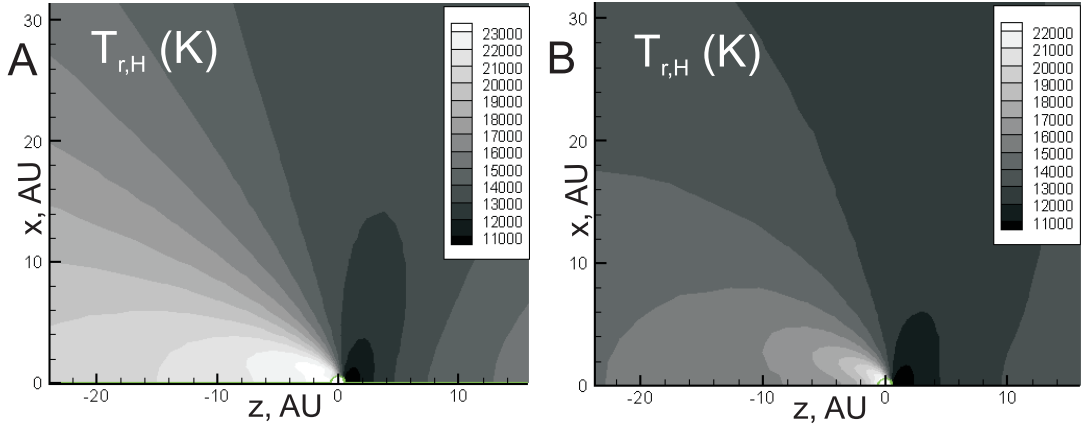}
\end{center}
\caption{The kinetic radial temperature ($T_{r}$) for the sum of primary and secondary interstellar atoms in the heliosphere; Plot A
corresponds to the two-component hot model 2, plot B corresponds to the results of the model 3. In these calculations $\mu=1$, $\beta_{E}=5.9 \cdot 10{-7}\, s^{-1}.$ }
\label{Tr_H}
\end{figure}

Let us return to model 3 without any additional assumptions.
It is seen (curve~3) that there is no minimum of $T_{los}$ at 50-60$^{\circ}$ at all in this case. But there is a small maximum at $\theta=150^{\circ}$.
This maximum can be explained by $\theta$-dependence of the kinetic temperature $T_{r}$. The radial temperature
of sum of the primary and secondary interstellar atoms is shown in Fig.~\ref{Tr_H} (plot A corresponds to model 2, plot B -- to model 3). It is seen that
in case of the two-component hot model 2 ionization leads to maximum of the $T_{r}$ in downwind. For model 3 the maximum of $T_{r}$
is located at about $\theta=150^{\circ}$. This effect is reflected in the line-widths of Ly-$\alpha$ radiation that is seen in curve 3 of Fig.~\ref{T_tot}~B.

\section{Conclusions and discussion}

In this work we applied different models of hydrogen distribution in the heliosphere to compute the spectral properties of the
backscattered solar Ly-$\alpha$ radiation as it would be measured at 1 AU in the anti-solar directions. We have found out that imprints of the heliospheric interface in the H atom distribution inside the heliosphere have
a strong influence on the Ly-$\alpha$ parameters.

We considered the minimum of the line-width of the backscattered Ly-$\alpha$ radiation at 50-60$^{\circ}$ from upwind that was observed by SWAN \citep{costa99, quem_etal06}. In the experimental data the line-width in the directions of $\theta=$50-60$^{\circ}$ is smaller than in the upwind direction by 1500-2000 K. This minimum is seen for 1996 and 1997 and practically is not seen for 2002-2003 years although data points for small angles are absent for this period \citep[see][Fig.~7]{quem_etal06}. This minimum was explained \citep[see][]{costa99, quem_etal06} by existence of the two different populations of the interstellar hydrogen
atoms that are shifted in velocity space. However, we noticed that the line-width calculated with 2D stationary Baranov-Malama model
 \citep{quem_izmod02} has no any minimum of T$_{los}$. Non-stationary 2D Baranov-Malama model \citep{quem_etal08} predicts small minimum in 2003, but there is no minimum in 1997.

In this work we theoretically explore the nature of the observed minimum on the basis of three types of the hydrogen distribution inside the heliosphere.
It was shown that the minimum of the line-widths appears only for the two-component hot model, and there is no minimum at all for model 3 that
takes into account all effects of the heliospheric interface.
It was found that the absence of the minimum in model 3 is because the effect of the two components is compensated by the non-Maxwellian features of the velocity distribution of H atoms at 90 AU after they passed the heliospheric interface region, namely,
by strong anisotropy of the kinetic temperatures of H atoms ($T_{z}<T_{\rho}$).

Therefore, the question why the
the minimum of the Ly-$\alpha$ line-width exists in the experimental data remains to be open. Possibilities to get the minimum still exist in the frame of model 3.
Firstly, models considered here does not take into account the effects of latitudinal and solar cycle variations of the photoionzation and charge exchange rates as well as the solar radiation pressure. These local effects may potentially change the result of this paper.

Another possibility is to change the boundary conditions at 90 AU, i.e. to make a change in the model of the heliospheric interface. For example, interstellar magnetic field may play a key role \citep{izmod_alex05}. Another possibility is the multi-component nature of both the heliospheric and interstellar plasmas \citep{malama_etal06, izmod_etal09, chalov_etal10}. In this non-equilibrium plasma  model the interstellar pickup ions are treated as the separate kinetic component. The plasma temperature in the vicinity of the heliopause is smaller in this model as compared with the Baranov-Malama model. Therefore, we could  expect a decrease in the kinetic temperatures of the secondary interstellar atoms. This might make larger velocity space separation of the primary and secondary interstellar atoms at 90 AU. The separation may effectively increase the effect of the two populations as compared with the other effects. As the result one may hope that the observed minimum will appear in the model.

All these possibilities will be checked in the nearest future.

This work was partially supported by RFBR grants 10-02-93113, 10-02-01316 and ``Dynastia'' foundation. Part of this work was done in the frame of ``FONDUE'' working group of International Space Science Institute in Bern.

\clearpage

\end{document}